# Causal resilience curves: A data-driven framework for quantifying the spatiotemporal impacts of metro service disruptions

Nan Zhang[a]
nan.zhang16@imperial.ac.uk

Daniel Hörcher[a]
d.horcher@imperial.ac.uk

Prateek Bansal[b]
prateekb@nus.edu.sg

Daniel J. Graham[a]*
d.j.graham@imperial.ac.uk

[a] Transport Strategy Centre, Department of Civil and Environmental Engineering, Imperial College London, London, UK

[b] Department of Civil and Environmental Engineering, National University of Singapore, Singapore

* Corresponding author

**Abstract**

Urban metro systems move vast numbers of passengers with a high level of efficiency in resource use, but frequently experience disruptions that result in delays, crowding, and deterioration in passenger satisfaction and patronage. To quantify these adverse consequences, this paper presents a novel, data-driven causal inference framework to measure metro resilience by estimating both the direct and spillover effects of service disruptions on passenger demand, journey time, travel speed and on-board crowding. By integrating high-frequency smart card data into a synthetic control design, we use weighted non-disrupted days to construct unbiased counterfactuals, which resolves confounding factors and accurately captures disruption propagation across the network. The impact estimates are further translated into station-level causal resilience curves that reveal spatial heterogeneity in the temporal patterns of degradation and recovery across locations, providing metro operators with actionable insights for targeted interventions and resource allocation. A case study of the Hong Kong MTR demonstrates the framework's superiority over naïve typical-day comparisons and machine-learning benchmarks in delivering unbiased resilience curves. This paper is the first to derive causal estimates of dynamic metro resilience. This practical tool can be generalised to evaluate resilience in a broad range of public transport systems.

## 1. Introduction

Metro systems form an important component of mass public transport in cities, characterised by large capacity and high-frequency services that can deliver vast volumes of passengers to central locations in small windows of time. However, metros experience disruptions frequently due to infrastructure or rolling stock failure, extreme demand shocks, bad weather or natural disasters, leading to a decline in service quality and ultimately in attractiveness and patronage [1]. Such disruption-induced performance losses are captured by the notion of metro operational resilience, that is the ability of an urban rail transit network to withstand shocks, absorb disturbances, sustain an acceptable level of service and restore full functionality within a tolerable timeframe. To manage and mitigate the adverse impacts, operators require accurate and unbiased evidence on how interruptions propagate through the network and affect different aspects of the passenger experience. Such knowledge forms the foundation for effective recovery strategies, future disruption management and providing real-time updates to passengers. Accordingly, understanding and enhancing metro resilience has become a core theme in contemporary public transport research [2].

Metro resilience curves provide a time-based visualisation of system performance changes over the full disruption life cycle: the steady pre-event state, the rapid degradation during the shock, and the staged recovery afterwards. Fig. 1 schematically illustrates the three phases. This dynamic profile evolved from the "resilience triangle" to more flexible, non-linear representations [3,4]. The curve links practical operational questions to quantitative metrics, for example, how deep is the performance loss? (vulnerability) and how quickly does service rebound? (rapidity). Operators and planners use these curves as decision support tools to rank critical network elements, schedule repair crews, pre-position spare trains and test alternative recovery plans [5-7]. Comparing the area under competing curves also supports cross-city benchmarking of metro resilience under various incident types [8]. The insight offered by such visualisations, however, is only as reliable as the underlying disruption impact estimates. Inaccurate quantification of resilience curves can mislead both disruption management and long-term investment decisions.

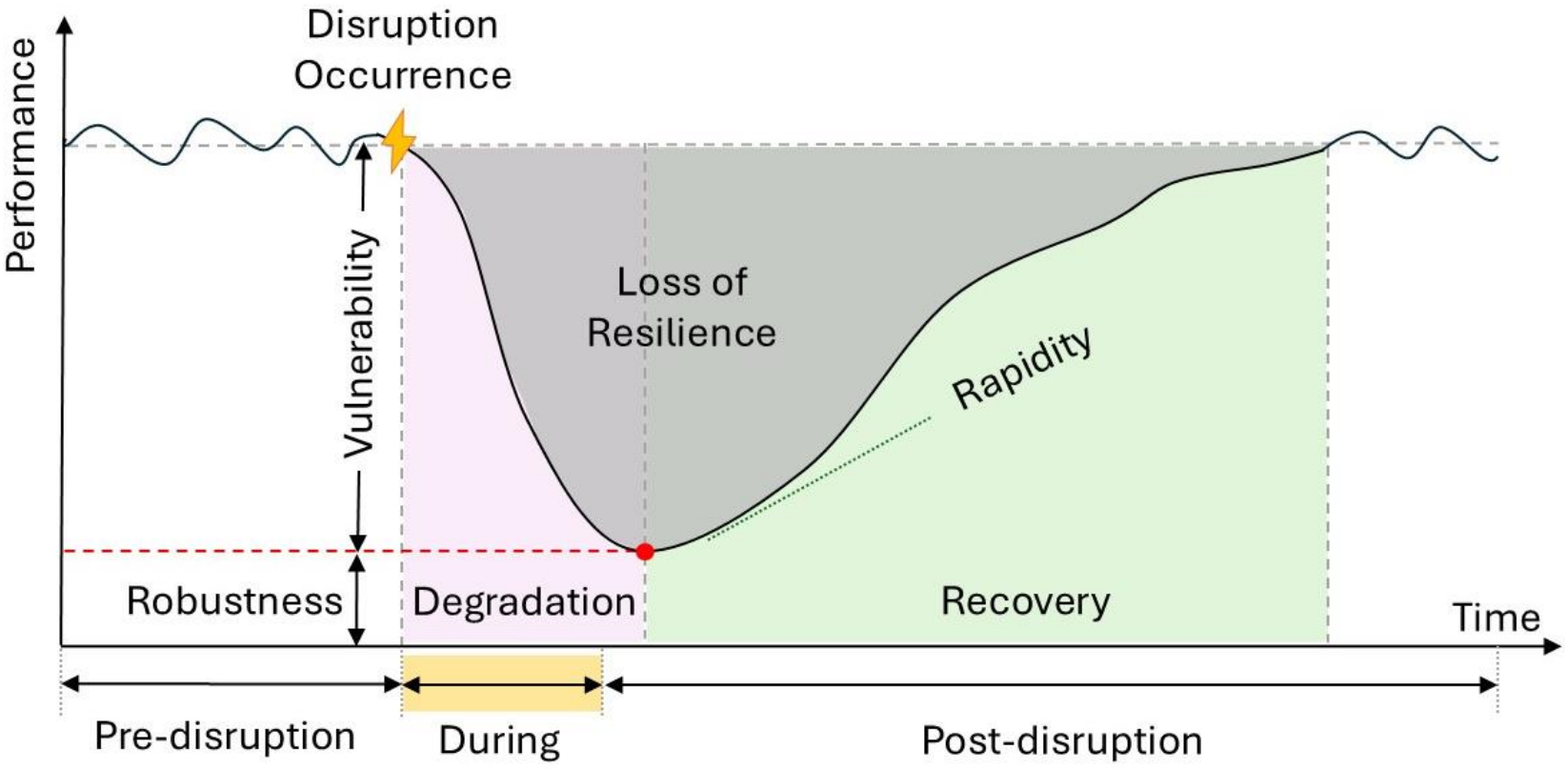

**Fig. 1.** Schematic of the dynamic resilience of urban metro systems.

Recent studies derive metro resilience curves primarily through two methodological stands, each with inherent limitations. Simulation-based approaches construct hypothetical disruption scenarios, while baseline service performance is known, the performance under the simulated interruption (unobservable counterfactual) must be inferred [9]. This inference depends on behavioural assumptions about passenger responses and network interactions, which are often over simplified or unrealistic, leaving the resulting curves highly sensitive to modelling choices [10,11]. Data-driven approaches, by contrast, exploit real incident records. Their disrupted performance is directly observed, while how the system would have behaved had the disruption not occurred (the counterfactual) must still be approximated, typically by reference to "normal" operating days [12-14]. This substitution is acceptable only if failures occur randomly, allowing one to credibly assume that the counterfactual performance would match the regular operating scenario, but this assumption is rarely met. Factors such as time of day, signalling type, passenger demand, and adverse weather simultaneously affect both failure likelihood [15-17] and passenger behaviour [18,19], illustrated in Fig. 2. These confounders bias resilience estimates when normal-day performance is used directly as a benchmark (ignoring the different distributions of confounding factors). Mis-quantified resilience curves, in turn, misinform disruption management and capital planning decisions, diverting resources from true bottleneck and compromising the objective of genuine resilience improvement.

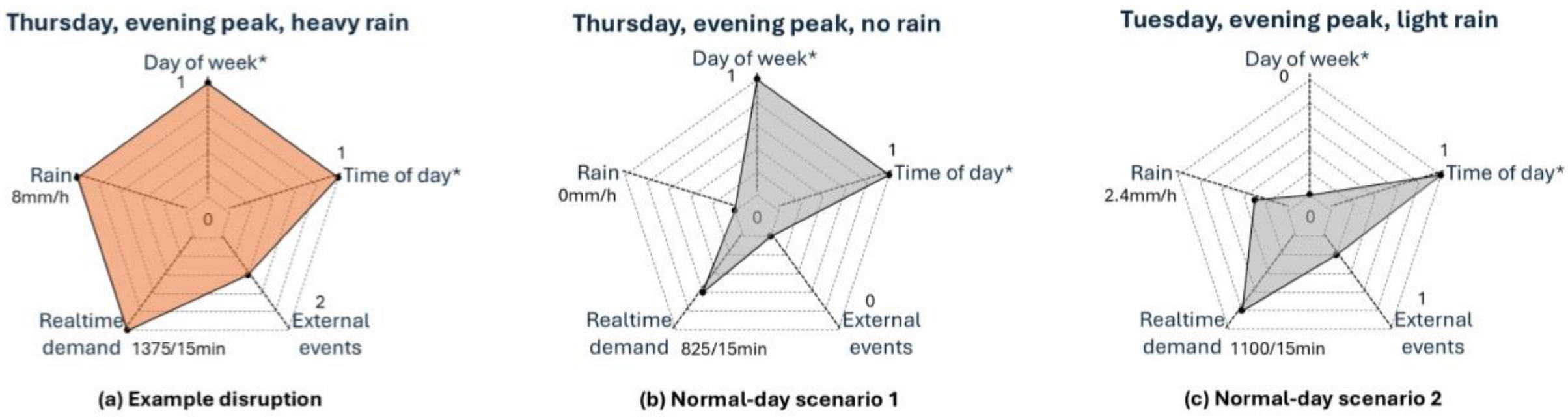

(a) Example disruption (b) Normal-day scenario 1 (c) Normal-day scenario 2

**Fig. 2.** The distribution of confounding factors for an example disruption and the two normal-day baseline scenarios. **Day of week* and *Time of day* are dummy variables.

To address the above gaps, this paper develops a causal inference framework for station-level resilience quantification using historical disruption data, where the performance measures of interest are passenger demand, average travel speed/journey time, and on-board crowding level. By utilising multi-day high-frequency smart card data (over 4.85 million trips per weekday), we adapt the synthetic control method to construct "no-disruption" counterfactual via *weighted* average of days without incidents anywhere in the network. By relaxing the strict non-interference assumptions[1] that is common in standard causal analyses, our framework captures how disruptions propagate through connected links. This design avoids many of the biases arising from the confounding issues while also tackling spillover impacts. A case study using data from Hong Kong Mass Transit Railway (MTR) demonstrates the practical insights this method delivers. The empirical evidence revealed in this study suggests that the synthetic control approach achieves unbiased resilience curves, outperforming both the normal-day comparisons and the machine-learning based predictions.

This study advances the field of reliability engineering and metro resilience in four ways. First, we propose a novel data-driven framework that causally quantifies the resilience of metro networks while explicitly correcting for confounding bias associated with non-random disruptions. Second, we introduce empirical causal resilience curves, as visual tools that integrate vulnerability, robustness, and recoverability at the station level and can be aggregated to form a network-wide picture. Third, we rigorously model network spillover effects, revealing heterogenous resilience trajectories across locations and quantifying how disruptions propagate. Fourth, by leveraging the unique scale and granularity of automated metro data, we enable event-specific causal analyses of system resilience that enrich the insights from crude aggregate averages. Collectively, these contributions widen the theoretical and practical foundations of metro resilience research, moving the literature beyond traditional simulation-based and naive data-driven approaches.

[1] Refers to the Stable Unit Treatment Value Assumption (SUTVA), which states each unit's outcome is solely determined by its own treatment and is unaffected by how other units are treated.

The remainder of the paper is organised as follows. Section 2 reviews current work on metro resilience quantification and resilience curve metrics. Section 3 details the causal inference foundations and the modified synthetic control framework. Section 4 describes the Hong Kong MTR case study, data sources and evaluation design. Section 5 presents our main empirical findings, including performance tests, disruption spillover patterns, causal resilience curves, and discusses future research directions. Section 6 concludes, highlighting the framework's potential to strengthen operational resilience in urban metro systems.

## 2. Literature review

Extensive research has focused on the resilience of urban metro networks, with comprehensive reviews of these studies provided by Wei et al. [9] and Hu et al. [20]. More broadly, transport system resilience has been surveyed by Mattsson and Jenelius [21], Wan et al. [22], Zhou et al. [23], and Bešinović [24]. Building on these works, we adopt the following updated definition of resilience: the ability of a metro system to withstand shocks, absorb disturbances, sustain an acceptable level of service and restore full functionality within a tolerable timeframe. This definition acknowledges the existence of different phases in the full life cycle of a disruption, intending to reflect the dynamic performance changes over time.

### 2.1 Quantification of metro resilience

The literature is rich in employing simulation models to investigate how hypothetical disruption scenarios affect metro performance [9,25,26]. Early resilience studies rely on topology and complex network theory, representing the metro network as a scale-free graph and measuring structural change when nodes or links are removed from the network [21,27,28]. Classic metrics such as node importance, betweenness centrality, and global efficiency reveal how connectivity degrades under random failures or targeted attacks. Beijing [29], London [30], Shanghai [31], Guangzhou [32], Zhengzhou [33], and more systems from other cities [34] have been evaluated in this way.

Based on pure topological analyses, more advanced resilience studies embed operational detail into the simulations, coupling passenger assignment process with network accessibility, timetable, and train scheduling constraints [35]. Within these models, disruption impacts are quantified through changes in demand loss, ridership distribution, passenger delay, operating cost, and crowding under experimental interruption settings [36-43]. For instance, D'Lima and Medda [44] used stochastic passenger counts to estimate resilience, while Sun et al. [45] derived vulnerability indices from platform and onboard passenger flow data. Recognising that route choices often need to change during disruptions, Sun and Guan [46] introduced passenger betweenness centrality and missed trip metrics, and Yin et al. [47] generalised flow-weighted betweenness to station, link and line closures. To capture the spatial propagation of disruptions, Shelat and Cats [48] combined stochastic user equilibrium assignment with link criticality scores. Chen et al. [49] incorporated stated travel preferences into effective path betweenness measures and Sun et al. [50] simulated cumulative affected node flows. Many of other investigations rely on BusMezzo, a mesoscopic public transport assignment platform that dynamically simulates individual route choices [51,52]. Using BusMezzo, Cats and Jenelius [53] quantified short-horizon and unplanned incidents in terms of passenger welfare and rolling stock costs, while Malandri et al. [54] estimated changes in the volume-capacity ratio to display network crowding spillovers.

Simulation approaches offer two clear advantages: (i) they do not require incident data, and (ii) they allow practitioners to test a wider range of scenarios, from single station or link closure to network collapse [30-32,37,38,46,55]. However, researchers need to make behavioural assumptions to infer passengers' response to virtual disruptions, which may not hold even if they are derived from patterns in observational data. The uncertainty in passengers' responses when facing an actual incident affects the validity of such assumptions. For example, many studies assume that all travellers have identical walking speeds, or they do not change destinations during disruptions unless there is no available route [36,37]. By contrast, field evidence shows that passengers typically travel at different speeds, they do reroute and change their destinations, or entirely leave the metro system even if a feasible path still exists to their original destination. Such modelling misspecification propagates into biased performance estimates and, ultimately, misleading resilience curves.

In view of the above concerns, empirical research has gained increasing attention, supported by growing access to a widening range of data sources; from user surveys [56-58] to large-scale automated records such as smart card data and train movement data. The latter have emerged as the mainstream because they offer high temporal accuracy, low data collection cost, and the possibility of long-term observations [59,60]. Using smart card and real incident data, a common strategy of assessing metro resilience is to contrast system performance on incident days with those on "typical" days. Sun et al. [61] estimated the total delay effects on three groups of travellers via alternations in passenger assignment outcomes. Chan and Schofer [62] evaluated New York City's Subway resilience to severe weather via variations in revenue vehicle mileage. Subsequent analyses have adopted similar designs to examine demand and journey time shifts [12], ridership under extreme rainfall [13], and tap-in reductions at affected stations [14]. At the individual level, Mo et al. [63,64] developed a probabilistic framework that infers traveller responses to unplanned incidents in Chicago's tap-in-only system. Unlike previous studies, Yin et al. [65] trained a Bayesian network on historical failure records from the Beijing Subway to investigate the relationship between system resilience and different incident categories.

Most of these empirical studies, however, treat metro disruptions as if they occur randomly, thereby ignoring the existence of confouding factors that affect both the occurrence of disruptions and their consequences. As illustrated in Fig. 2, failure risk is systematically higher at peak hours, under heavy demand, or during adverse weather [10,15-17], and these same factors also magnify their impact on disrupted performances [18,19]. Directly comparing incident-day outcomes with normal-day baselines therefore yields biased resilience estimates and, by extension, unreliable resilience curves [66].

A related literature employs predictive models trained on past incidents to forecast future disruption impacts. Silva et al. [67] predicted the exit ridership and passenger behaviour for unseen scenarios, such as station closure and line segment closure. Yap and Cats [68] applied supervised learning approaches to predict disruption exposure and passenger delays caused by it. Zhao et al. [69] developed two representative tree-based methods and a deep learning-based model to predict the ridership affected by unplanned incidents. Liu et al. [70] proposed a multiple linear regression model to predict the duration of disruption impact on passenger trips. Although operationally useful, these forecasts remain associational. They are unable to quantify causal effects because they disregard the root problem of confouding.

One recent attempt to address this bias is the propensity score matching (PSM) framework of Zhang et al. [10], which relaxes the random-disruption assumption by balancing the internal and external confounding factors. They estimated the average causal effects of historical incidents on disrupted stations. Yet the conventional PSM method cannot accommodate spillover effects (also known as the "interference" phenomenon); that is, the possibility that a failure at one station may influence service quality at neighbouring and even relatively distant (connected or adjacent) stations, violating the assumption that disruptions affect station performance independently. The PSM method cannot deliver event-specific estimates either, as it targets the average causal effect across many incidents [71]. These limitations motivate our pursuit for new empirical tools that address confounding, model impact propagation through the network, and reveal disruption-specific causal impacts simultaneously.

**Table 1.** Summary of metro resilience studies: approaches and key features.

| | Publication | Target system | Network topology | Service operation | Incident data | Causal design | Disruption propagation |
|---|---|---|---|---|---|---|---|
| Simulation-based | Angeloudis & Fisk [27] | Multiple | √ | | | | |
| | Derrible & Kennedy [28] | Multiple | √ | | | | |
| | Yang et al. [29] | Beijing Subway | √ | | | | |
| | Chopra et al. [30] | London Underground | √ | | | | |
| | Yang et al. [72] | Beijing Subway | √ | | | | |
| | Wang et al. [34] | Multiple | √ | | | | |
| | Zhang et al. [31] | Shanghai Metro | √ | | | | |
| | Zhang et al. [32] | Guangzhou Metro | √ | | | | |
| | Qi et al. [33] | Zhengzhou Metro | √ | | | | |
| | Rodríguez-Núñez & García-Palomares [36] | Metro Madrid | | √ | | | |
| | D'Lima & Medda [44] | London Underground | | √ | | | |
| | Sun et al. [45] | Shanghai Metro | √ | √ | | | |
| | Adjetey-Bahun et al. [37] | Paris Mass Railway | | √ | | | |
| | Sun & Guan [46] | Shanghai Metro | √ | √ | | | |
| | Yin et al. [47] | Beijing Subway | √ | √ | | | |
| | M'cleod et al. [35] | New York City Subway | | √ | | | |
| | Shelat & Cats [48] | Amsterdam Metro | √ | √ | | | √ |
| | Cats & Jenelius [38] | Stockholm Metro | | √ | | | |
| | Lu [39] | Shanghai Metro | √ | √ | | | |
| | Malandri et al. [54] | Stockholm Metro | | √ | | | √ |
| | Sun et al. [50] | Beijing Subway | √ | √ | | | |
| | Nian et al. [40] | Shanghai Metro | √ | √ | | | |
| | Xu et al., [41] | Multiple | √ | √ | | | |
| | Chen et al. [49] | Chengdu Subway | √ | √ | | | |
| | Ma et al. [42] | Beijing Subway | √ | √ | | | |
| | Xu & Chopra [43] | Hong Kong MTR | √ | √ | | | |

| | | | | | | |
|---|---|---|---|---|---|---|
| Prediction-based | Silva et al. [67] | London Underground | √ | √ | | |
| | Yap & Cats [68] | Washington Metro | √ | √ | | |
| | Zhao et al. [69] | Anonymous | √ | √ | | |
| | Liu et al. [70] | Anonymous | √ | √ | | |
| Empirical | Chan & Schofer [62] | New York City Subway | √ | √ | | |
| | Sun et al. [61] | Beijing Subway | √ | √ | | |
| | Liu et al. [12] | Anonymous | √ | √ | | |
| | Zhang et al. [10] | London Underground | √ | √ | √ | |
| | Zhou et al. [13] | Shenzhen Metro | √ | √ | | √ |
| | Mo et al. [63,64] | Chicago 'L' | √ | √ | | |
| | Yin et al. [65] | Beijing Subway | √ | √ | | |
| | Zhou et al. [14] | Beijing Subway | √ | √ | | √ |
| | **This paper** | Hong Kong MTR | √ | √ | √ | √ |

## 2.2 Resilience curve metrics

After a resilience curve has been defined, the core challenge is to quantify its shape, that is to characterise how system performance evolves over time. In the absence of a closed form expression for the curve, a practical approach is to design a set of scalar metrics that summarise the curve's key features and performance dynamics throughout the disruption [11]. Although terminology varies across disciplines, in the engineering context these metrics mainly fall into four categories: magnitude, duration, integral, and rate [4,73].

*Magnitude-based metrics* describe performance at specific time points. Typical examples include maximum performance loss (depth of impact), residual performance (minimum functionality reached), and restored performance or degree of recovery [4]. *Duration-based metrics* measure the temporal span between key milestones, such as degradation time (onset to nadir) and recovery time between nadir and full or partial restoration [4]. *Integral-based metrics* combine both time and magnitude, most commonly the loss of resilience, the area between the observed curve and the pre-disruption status [74]. This index reflects the cumulative performance loss experienced by travellers. *Rate-based metrics* are obtained as the first derivative of the curve, such as the failure rate and recovery rate. These gradients indicate how rapidly the system loses or regains functionality and are often interpreted as proxies for adaptive capacity, resistance, and recovery efficiency [75]. All four categories have been widely adopted in recent metro resilience studies to benchmark networks, evaluate intervention strategies, and compare incident types [9,13,39,42,76]. Their joint use enables a multidimensional assessment that reflects the complex reality of service degradation and recovery.

## 2.3 Research gaps

Despite the substantial progress outlined above, three critical gaps remain in the metro resilience literature. First, simulation studies, even those embedding sophisticated assignment or agent-based modules, still rely on behavioural assumptions whose validity is rarely tested against real incidents. Their outputs therefore may be sensitive to behavioural assumptions and misrepresent the actual system response.

Second, most empirical analyses benchmark incident-day outcomes with "typical" days or train predictive models on historical disruptions. Both approaches ignore the issue of confounding stemming from non-random metro disruptions. Failing to adjust for confounding yields biased resilience curves and flawed decisions based on them.

Third, and more importantly from a network management perspective, existing approaches rarely reveal the dynamic propagation of disruptions and the spatial heterogeneity of resilience curves. They overlook how performance degrades and recovers differently across stations, and how cascading spillover effects reshape those trajectories. Without methods that capture this location-specific evolution, operators lack the granular insight needed to prioritise interventions where they matter most.

Collectively, these gaps call for a data-driven causal framework that (i) validates disruption effects through real-world observations; (ii) constructs credible counterfactuals to eliminate confounding bias; and (iii) traces the spatiotemporal propagation of disruptions, thus revealing heterogeneous resilience patterns across the entire metro network.

## 3. Methodology

In this section, we first introduce the key concepts and assumptions in causal inference, and then clarify the goal of our work. A data-driven and customised synthetic control framework is proposed. We present this specialised causal inference design in Section 3.2, which addresses the interference issues within the metro network. In Section 3.3, we mathematically model the construction of effective synthetic counterfactuals. Section 3.4 outlines how the estimated impacts are transformed into causal resilience curves.

### 3.1 Preliminaries in causal inference

To establish causality behind the treatment (or intervention) applied to a study unit, in our study a disruption, Rubin's potential outcomes framework is a foundational approach [77]. For unit $i$, let $W_i$ indicate the treatment assignment, and $Y_i$ denote the outcomes of interest. The potential outcomes for a binary treatment are defined as:

$$Y_i(W_i) = Y_i(0) \times (1 - W_i) + Y_i(1) \times W_i, \quad [1]$$

$Y_i(0)$ denotes the outcomes that would be attained if unit $i$ did not receive the treatment ($W_i = 0$). Conversely, $Y_i(1)$ denotes the outcomes that unit $i$ would attain if it was exposed to the treatment ($W_i = 1$) [71]. The individual treatment effect (ITE) is determined by comparing these two potential outcomes at the unit level, expressed as $Y_i(1) - Y_i(0)$. However, one of the two potential outcomes is inherently counterfactual and thus only a single outcome will be ultimately observed, which becomes a major challenge for ITE estimation.

The second ingredient of the Rubin causal model is the assignment mechanism, which is assumed non-random and defined as the conditional probability of receiving the treatment given a set of unit-specific background attributes $X_i = [X_{i1}, X_{i2}, \ldots, X_{iz}]^T$, where $z$ is the dimension of the attributes [66]. In observational studies, where the treatment assignment is non-random, three critical assumptions are required for valid estimation of causal effects [78].

i. Ignorability (or Unconfoundedness) Assumption:

Given a set of the covariates $X_i$ , the treatment assignment is independent of the potential outcomes, $W_i \perp\!\!\!\perp \left(Y_i(0), Y_i(1)\right) | X_i$ . It assumes that all the confounders are observed and measured.

ii. Positivity (or Overlap) Assumption:
Every study unit has positive possibility of receiving each treatment condition, $0 < Pr\left(W_i | X_i, Y_i(0), Y_i(1)\right) < 1$. It also implies that the distributions of covariates overlap for the treatment and control groups.

iii. Stable Unit Treatment Value Assumption (SUTVA):
This assumption ensures that the treatment applied to one individual does not affect the outcomes of another individual. It also implies that the treatment is consistent across all subjects. As shown in Eq. (1) but also could be noted that $Y(W_i, W_j) = Y(W_i)$ for all $j$.

However, in the context of metro networks, adjacent stations are connected by tracks and continuous train services. When one disruption occurs in a station, the adverse impacts such as delays and crowding can spread to the entire network via metro lines. The presence of interference among stations implies that the SUTVA is no longer plausible for disruption impact quantification. Thus, the goal of this study is to develop a novel causal inference framework that relaxes the SUTVA, and more importantly, leverages the unique structure of large-scale automated metro data to assess the spatiotemporal propagation of disruption impacts.

**3.2 Customised synthetic control framework for metro networks**

In this research, we treat metro disruptions as 'treatments' and the objective of our analysis is to quantify the direct and indirect causal effect of treatments on 'outcomes' related to the quality of service provision. Specifically, we are interested in estimating station-level impacts on travel demand, journey times, travel speed of passengers, and crowding density on board. The detailed definition of each outcome measure is provided in the Appendix.

We define the study unit as the status of a metro station $a = 1, \dots, A$ on a given day $d = 1, \dots, D$, during interval $t = 1, \dots, T$. We consider 15-minute-long intervals. The station is classed as *treated* if it encounters a service interruption of at least five minutes in the 15-minute interval. The treatment assignment variable, denoted by $W_{adt} \in \{0,1\}$, records whether station $a$ has been exposed to disruptions during interval $t$ on day $d$. Under the assumption that there are no hidden versions of the treatment (consistency assumption), see [78], we use $Y_{adt}(W_{adt})$ to denote the potential outcomes of metro service provision, namely the total inflow and outflow of passengers, the average journey time, average travel speed, and the density of crowding. More specifically,

$$Y_{adt} = \begin{cases} Y_{adt}(0) & if\ W_{adt} = 0 \\ Y_{adt}(1) & if\ W_{adt} = 1, \end{cases} \qquad [2]$$

where $Y_{adt}(0)$ and $Y_{adt}(1)$ are counterfactual potential outcomes, only one of which is observed.

To create the synthetic counterfactual outcome, we create a donor pool from data observed on days when disruptions did not happen in the entire metro network: $\boldsymbol{d_N}$ is a set of such undisrupted days with cardinality $J$. This design of the donor pool benefits from the fact that high-frequency smart card data contain observations for all time intervals from multiple days. To quantify the impact of a disruption that starts at station $a_I$ on day $d_I$ at time $T_{IS}$ and ends at time $T_{IE}$, we construct a vector of outcomes $\boldsymbol{p} = \{\boldsymbol{p_1}, \boldsymbol{p_2}, \dots, \boldsymbol{p_A}\}$, where $\boldsymbol{p_a}$ is the two-dimensional vector of outcomes for station $a$ during time intervals $t = T_{IS}, \dots, T$ on the disrupted day $d_I$ and $J$ undisrupted days (i.e., $J + 1$ days). We assume that this disruption has no effect on outcomes before the treatment period $T_{IS}$. Conversely, after $T_{IS}$, all stations in the network can be affected by this disruption. Since we stack the data of the treated day followed by undisrupted days, $p_{ajt} = Y_{ad_It}(W_{ad_It})$ for $j = 1$ and $p_{ajt} = Y_{ad_jt}(W_{ad_jt})$ for $j = 2, \dots, J + 1$, $d_j \in \boldsymbol{d_N}$. Note that $W_{ad_It} = 1$ if $t \geq T_{IS}$ and $W_{ad_It} = 0$ otherwise.

For a specific time interval of a treated/affected station $a$, the counterfactual outcome is defined as a weighted average of the outcomes in the donor pool, where $\boldsymbol{C^a} = (c_2^a, \dots, c_{J+1}^a)'$ is a $J \times 1$ vector of non-negative weights that sum to one [79]. See the next subsection for the way we determine these weights. The synthetic control estimators of the counterfactual outcomes is:

$$\hat{Y}_{ad_It}^N = \sum_{j=2}^{J+1} c_j^a \cdot Y_{ad_jt}(0) \qquad t = T_{IS}, \dots, T, \tag{3}$$

while the causal effect of the treatment is estimated by

$$\hat{\tau}_{ad_It} = Y_{ad_It} - \hat{Y}_{ad_It}^N \qquad t = T_{IS}, \dots, T. \tag{4}$$

With the definitions above, during and after a given disruption, the direct causal effects on a treated station $a_I$ (service interrupted at such station) is derived as

$$\hat{\tau}_{a_Id_It} = Y_{a_Id_It}(1) - \sum_{j=2}^{J+1} c_j^{a_I} \cdot Y_{a_Id_jt}(0) \qquad t = T_{IS}, \dots, T. \tag{5}$$

where $Y_{a_Id_It}$ denotes the observed outcome of the treated unit on the disrupted day in interval $t$. Furthermore, $c_j^{a_I}$ denotes the weight of the $j^{th}$ day in the corresponding donor pool for station $a_I$, and $Y_{a_Id_jt}(0)$ denotes the observed outcomes for the same station-interval pair on the $j^{th}$ day.

Similarly, the indirect spillover causal effects of a disruption on the performance of other station $a_O$ ($a_O \in 1, \dots, A \setminus a_I$, normal service at such station) is derived as

$$\hat{\tau}_{a_Od_It} = Y_{a_Od_It}(1) - \sum_{j=2}^{J+1} c_j^{a_O} \cdot Y_{a_Od_jt}(0) \qquad t = T_{IS}, \dots, T, \tag{6}$$

where $Y_{a_Od_It}(1)$ denotes the observed outcomes for the affected units of other (non-disrupted) stations during and after a given disruption; ${c^{a_O}}_j$ and $Y_{a_Od_jt}(0)$ denote the weight and outcomes of the $j^{th}$ day in the corresponding donor pool for station $a_O$. Fig. 3 illustrates the design of the synthetic control framework for metro disruptions.

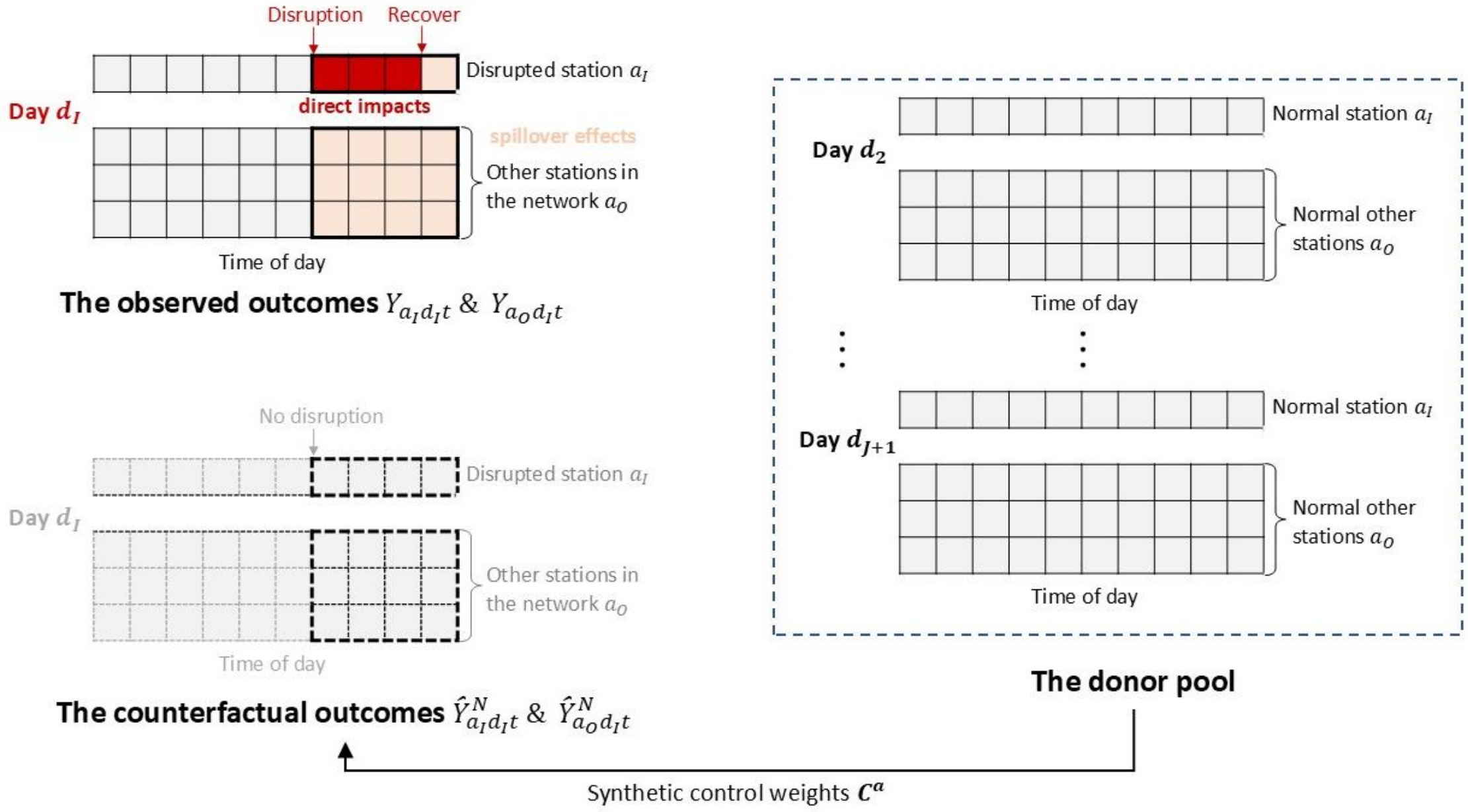

**Fig. 3.** Schematic overview of the modified synthetic control method for metro disruptions. The donor pool consists of observations from non-disrupted days, and $a_O$ represents any other undisrupted station in the network.

It is worth noting that standard synthetic control methods need to follow the SUTVA. Otherwise, post-treatment controls will be contaminated by spillover effects, resulting in a biased estimate of counterfactual potential outcomes. A key contribution of this work is the introduction of a smart and intuitive modification to the donor pool design, made possible by exploiting the high-frequency nature of automated metro data. By leveraging time series observations throughout multiple days (including both disrupted and normal days), all control units comprising the donor pool are selected exclusively from days without disruptions. That is, such an adapted donor pool would not be affected by any treatment, which therefore naturally relaxes the SUTVA and enables the unbiased estimation of direct and spillover causal effects. Moreover, this longitudinal-data-based synthetic control framework also facilitates the estimation of individual disruption effects.

### 3.3 The choice of weights

A simple way of constructing synthetic counterfactuals is to assign equal weights $c_j^a = 1/J$ to each unit in the donor pool. The estimator for $\tau_{ad_It}$ is then

$$\hat{\tau}_{ad_It} = Y_{ad_It} - \frac{1}{J}\sum_{j=2}^{J+1} Y_{ad_jt} \qquad t = T_{IS}, \dots, T, \qquad [7]$$

where the synthetic control is the unweighted average of observed historic outcomes in the donor pool.

In this research, we apply the method proposed by Abadie and Gardeazabal [80] and Abadie et al. [81,82] to determine $\boldsymbol{C^a}$. For the disrupted day $d_I$ and each day in the donor pool $d_j$ corresponding to station $a$ at time $t < T_{IS}$, we first collect data on a set of $k$ predictors[2] of the outcomes, denoted by

[2] Predictors refer to the set of pre-intervention variables used to forecast or explain the outcome of interest. These predictors can include lagged values of the outcome itself as well as other relevant covariates that capture underlying characteristics of the treated unit.

$k \times 1$ vectors $\boldsymbol{X_1^{at}},\ \boldsymbol{X_2^{at}}, \ldots, \boldsymbol{X_{J+1}^{at}}$. Let $\boldsymbol{X_1^a} = \left(\frac{\sum_{t \in T_0} \boldsymbol{X_{11}^{at}}}{|T_0|}, \ldots, \frac{\sum_{t \in T_0} \boldsymbol{X_{1k}^{at}}}{|T_0|}\right)$ be a $k \times 1$ vector and collect the values of such predictors at the disrupted day for a pre-intervention period $T_0 \subseteq \{1, 2, \ldots, T_{IS} - 1\}$. Similarly, the $k \times J$ matrix $\boldsymbol{X_0^a} = \left[\boldsymbol{X_2^a}, \ldots, \boldsymbol{X_{J+1}^a}\right]$ represents the predictors for the $J$ non-disrupted days within this donor pool. Predictors $\boldsymbol{X}$ are selected such that they are unaffected by the treatment (service interruption), but they do influence the outcomes, which may include pre-interruption values of $Y_{adt}$.

Weights $\boldsymbol{C^a}$ are optimised to ensure that the resulting synthetic control units best resemble all relevant characteristics (predictors) of the treated unit before the disruption. That is, given a set of non-negative constants $\boldsymbol{V^a} = (v_1^a, \ldots, v_k^a)$, the optimal synthetic control weight vector $\boldsymbol{C^{a^*}} = \left(c_2^{a*}, \ldots, c_{J+1}^{a*}\right)'$ is obtained from the following minimisation problem:

$$\min_{\boldsymbol{C^a}} \|\boldsymbol{X_1^a} - \boldsymbol{X_0^a} \cdot \boldsymbol{C^a}\|_{V^a} = \frac{1}{|T_0|} \sqrt{\sum_{h=1}^{k} v_h^a \cdot \sum_{t \in T_0} \left(X_{1h}^{at} - c_2^a \cdot X_{2h}^{at} - \cdots - c_{J+1}^a \cdot X_{(J+1)h}^{at}\right)^2},$$
$$such\ that\ \sum_{j=2}^{J+1} c_j^a = 1,\ c_j^a > 0, \qquad [8]$$

where the positive constants $v_1^a, \ldots, v_k^a$ prioritise the $k$ predictors by assigning different levels of importance to each covariate.[3] Each potential choice of $\boldsymbol{V^a}$ produces a corresponding set of synthetic control weights $\boldsymbol{C}(\boldsymbol{V^a}) = \left(c_2^a(\boldsymbol{V^a}), \ldots, c_{J+1}^a(\boldsymbol{V^a})\right)'$. We choose $\boldsymbol{V^a}$, such that $\boldsymbol{C}(\boldsymbol{V^a})$ minimises the mean squared prediction error (MSPE) of this synthetic control with respect to outcome $Y_{ad_I t}^N$ before the disruption:

$$\min_{\boldsymbol{V^a}} \sum_{t \in T_0'} \left(Y_{ad_I t} - c_2^a(\boldsymbol{V^a}) \cdot Y_{ad_2 t} - \cdots - c_{J+1}^a\,(\boldsymbol{V^a}) \cdot Y_{ad_{J+1} t}\right)^2,$$
$$such\ that\ \sum_{h=1}^{k} v_h^a = 1,\ v_h^a > 0, \qquad [9]$$

where the synthetic control weights $c_2^a(\boldsymbol{V^a}), \ldots, c_{J+1}^a(\boldsymbol{V^a})$ are functions of $\boldsymbol{V^a}$, for a pre-intervention period $T_0' \subseteq \{1, 2, \ldots, T_{IS} - 1\}$, $T_0' \neq T_0$.

To determine the optimal values of $\boldsymbol{V^a}$ and $\boldsymbol{C^a}$, inspired by Abadie [79], we present the detailed steps in Algorithm 1.

---

**Algorithm 1** Synthetic control weights optimisation

---

**Input**: Observed outcomes $Y_{adt}$ and predictors $\boldsymbol{X^a}$ in pre-treatment periods
**Output**: Optimal values of $\boldsymbol{V^a}$ and $\boldsymbol{C^a}$

1. Initialise training period ($t = 1, \ldots, t_0$) and subsequent validation period ($t = t_0 + 1, \ldots, T_{IS} - 1$), by dividing the pre-disruption periods
2. Use training period data on $\boldsymbol{X^a}$, obtain $\boldsymbol{\widetilde{C}^a}(\boldsymbol{V^a})$ by solving the optimisation problem in Eq. [8]
3. Use validation period data on $Y_{adt}$, obtain the optimal $\boldsymbol{V^{a*}}$ by solving Eq. [9]
4. Use validation period data on $\boldsymbol{X^a}$ and the resulting $\boldsymbol{V^{a*}}$, obtain the final $\boldsymbol{\widetilde{C}^{a^*}} = \boldsymbol{\widetilde{C}^a}(\boldsymbol{V^{a^*}})$ by solving Eq. [8]

---

[3] $\boldsymbol{V^a}$ measure the distance between the treated unit's characteristics and those of the control units, which are tuning parameters for predictor relevance. Both $\boldsymbol{V^a}$ and $\boldsymbol{C^a}$ are learnable weights from the pre-treatment data.

### 3.4 Innovations in dynamic resilience visualisation

A key output of our framework is the station-level time series of disruption impacts for five outcome variables: (i) entry ridership, (ii) exit ridership, (iii) average journey time, (iv) average travel speed, and (v) crowding density. These causal effect series can be directly converted into *causal resilience curves*, providing a clear visual and quantitative description of how service performance at each station deteriorates and subsequently recovers from a disruption in real time. Below, we outline the steps to construct such curves.

**Define a performance measure**: we first transform the impact measures $\hat{\tau}_{ad_I t}$ into *resilience-oriented* performance metrics. Specifically:

- Demand loss: reduction in entry and exit ridership
- Passenger travel speed loss[4]: decrease in average travel speed
- Passenger comfort loss: $-1 \times$ (increase in onboard crowding density)

**Plot the dynamic curve**: a causal resilience curve for station $a$ is simply the chosen performance measure plotted against time $t$. In practice, metro operators may wish to summarise the typical shape of the curve and extract key resilience metrics. These metrics can be calculated separately for each outcome variable to form a multidimensional resilience profile of how demand, travel speed, and crowding recover.

- Magnitude of performance loss ($max\left|\hat{\tau}_{ad_I t}\right|$): the greatest performance gap relative to the undisrupted counterfactual.
- Area of performance loss ($\sum_t \left|\hat{\tau}_{ad_I t}\right|$): an aggregated measure capturing both the severity and duration of the disruption, referring the overall loss of resilience.
- Degradation time ($T_{deg}$) and recovery time ($T_{rec}$): the duration of the performance metric to reach the maximum loss, and the duration from the maximum loss to revert toward pre-disruption levels.
- Failure and recovery rates: the first derivative of the curve, indicating how rapidly the system loses or regains functionality. Different combinations of these rates determine whether the curve is concave or convex during the degradation or recovery phase.

**Extend across the network**: by constructing dynamic resilience curves for all stations, we can build a complete picture of how disruptions propagate and resolve throughout the network. In particular, identifying stations with prolonged decreases in ridership, speed, or excess crowding highlights potential system bottlenecks and indicates where targeted interventions may accelerate recovery.

## 4. The Case study

We conduct a case study to evaluate the customised synthetic control on real-world metro system datasets, to answer the following questions: (1) Empirically, is the proposed framework effective for operators to assess the spatiotemporal effects of metro disruptions? (2) Does our approach provide a more accurate quantification of resilience compared to existing methods?

---

[4] Considering that changes in average journey time may include variations caused by altered destination choices (travel distance changes) during disruptions, we use average travel speed to construct resilience metrics.

### 4.1 Data

This case study utilises large-scale automated data from four urban lines of Hong Kong MTR, the Island Line, Tsuen Wan Line, Kwun Tong Line and Tseung Kwan O Line, with 49 stations in total. A map of the partial network that we study is provided in Fig. 4. The following data are used to estimate the direct and spillover causal effects of disruptions.

*Pseudonymised smart card data*: The Hong Kong MTR provided smart card data from 01/01/2019 to 31/03/2019 (over 4.85 million trips per day). The dataset contains information on the time and location of tap-in and tap-out transactions throughout the system, recording individual trips. Based on the data, we compute aggregate passenger flows at station entries and exits, passenger's average journey time, the average travel speed [10], and crowding density [83] for each target station. The resolution of time stamps exacts to one second.

*Automated vehicle location (AVL) data and incidents logs:* The MTR provided AVL data and incident information data during the same study period, which are used to generate historical disruption logs [84]. The AVL data contain information on train ID, service ID, the timestamp of train movements (including precise departure and arrival times), and the location of train movements (including station, line and directions). The resolution of time stamps is exact to one second. Incident logs are manual inspection record of incidents, including information on the time and location, cause and duration of disruptions. Readers are referred to Appendix for more details on our disruption data.

*Weather data:* We collect data on outside temperature, wind speed and precipitation status from the web portal Weather Underground of Hong Kong. Based on hourly historical observations, we estimate weather conditions for all selected stations at 15-minute intervals.

*Mega events in Hong Kong:* From 01/2019 to 03/2019, we collect information, including the location and time, on three types of mega-events held in Hong Kong: concerts, sports matches and exhibitions. Data sources include official news and government records.[5]

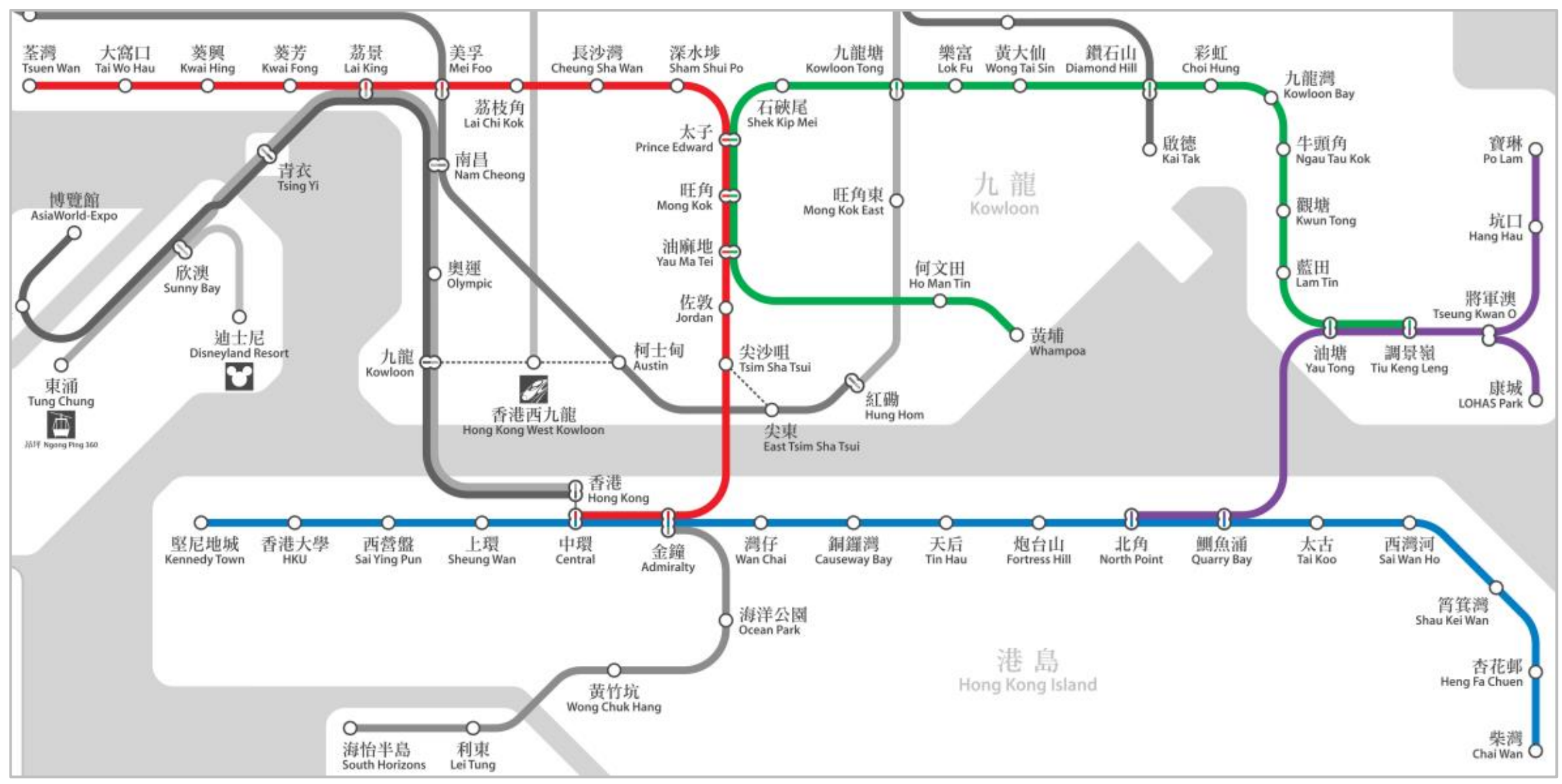

[5] https://www.mevents.org.hk/en/index.php.
https://www.lcsd.gov.hk/tc/programmes/programmeslist/mqme_prog.html.

**Fig. 4.** The map of four urban lines that we study in the MTR network (highlighted in colour).

### 4.2 Design and setup

Our study period covers 54 weekdays, of which 13 weekdays with no disruption are used to construct the donor pool. Within this period, we randomly selected three service disruptions, each occurring at a different type of station and at various times of day, to generate a diverse set of evaluation scenarios. Table 2 summarises the detailed information of the three disruptions. [6]

**Table 2.** Details of the three example disruptions.

| ID | Occurrence Location | Weekday | Occurrence Time | Duration |
|---|---|---|---|---|
| 1 | Terminal Station | Mon | 17:41 | 27min |
| 2 | Transfer Station | Wed | 20:31 | 12min |
| 3 | Regular Station | Fri | 17:47 | 6min |

The time of a service day is divided into 72 intervals of 15 minutes each, and the metro station in each 15-minute interval (station-interval) is our study unit. To account for the non-randomness of disruption occurrence, we consider partial confounding factors of metro disruptions when selecting predictors, such as weather conditions, day of the week and external events within the city. These are summarised in Table 3.

**Table 3.** Potential predictors of metro performance.

| Category | Predictors | Description |
|---|---|---|
| Pre-intervention outcomes (15-minutes) | Entry ridership | The number of passengers that enter the study unit before the disruption starts. |
| | Exit ridership | The number of passengers that exit the study unit before the disruption starts. |
| | Average journey time | The average journey time of passengers that enter the study unit before the disruption starts. |
| | Average travel speed | The average travel speed of passengers that enter the study unit before the disruption starts. |
| | Crowding density | The onboarding crowding level measured by the number of passengers per square meter. |
| Weekday | Day of week | Dummy variable, representing whether it is on the same day of the week as the disrupted date. |
| Weather conditions | Temperature | Atmospheric temperature around study units, ranging from 15℃ to 27℃. |
| | Wind speed | The wind speed around study units, ranging from 4 to 44 km/h. |
| | Rain status | Rain precipitation around study units, ranging from 0 to 4 mm/h. |
| External events | Concert | Dummy variable, indicating whether a concert is held in Hong Kong. Not considering its location within the city. |
| | Sports | Dummy variable, indicating whether a sports match is held in Hong Kong. Not considering its location within the city. |
| | Exhibition | Dummy variable, indicating whether a large-scale exhibition is held in Hong Kong. Not considering its location within the city. |
| | Overall mega-events | Dummy variable, indicating whether there are external mega-events held in Hong Kong. |

[6] Please note that each selected disruption is the only one that occurred on that day, which implies that there was no other disruption occurred across the entire network.

To estimate the direct and spatial-temporal spillover effects, for any existing station-interval pair, we create a corresponding synthetic control unit by weighting historic observations of the same station-interval pair from undisrupted days. The weights are set to maximise the synthetic control's ability to replicate observed exogenous characteristics (predictors) and metro performance outcomes in the immediate pre-intervention time intervals at the treated/affected station. For different performance measures, such as demand, travel speed, and crowding, etc., the process of generating the optimal synthetic control is totally independent for each station. In other words, under a specific disruption, any station within the network will have five distinct sets of weight combinations (both $\boldsymbol{V}$ and $\boldsymbol{C}$), corresponding to the five outcome indictors in this study. The causal inference framework and the computation process are implemented using $R$ and the relevant package '*Synth'*.

We compare our proposed method with the following competitive baselines. Model performance is benchmarked using the mean squared prediction error (MSPE) of the synthetic control with respect to factual outcomes before the disruption occurred.

- **Before-after comparison**: using the time-invariant average of pre-disruption observations.
- **Average control (AC)**: taking the unweighted average of the historic observations in the donor pool.
- **Single control**: comparing with a random control unit from the donor pool.
- **Linear regression** (LR)
- **Support vector machine** (SVM)
- **Random forest** (RF)
- **Extreme Gradient Boosting** (XGBoost)

The machine learning-based benchmarking models, including LR, SVM, RF, and XGBoost, are trained using data from donor pool units to develop predictive models for outcome measures, and their predictive performance is also tested using pre-treatment data from disrupted units.

## 5. Results and discussion

### 5.1 Synthetic control performance

Fig. 5 benchmarks the counterfactual predictive power of (i) our synthetic control design (black dashed line) against two baseline approaches: (ii) using the time-invariant average of pre-disruption observations (before-after comparison, blue dashed line) and (iii) taking the unweighted average of the historic observations of typical or normal days (average control, green dashed line). For each example disruption scenario, we first compare all three estimates to the pre-treatment period of the disrupted station. This figure shows that the naive before-and-after comparison cannot capture the changes in the pre-intervention time series of the outcome variables. Our weighted synthetic control can closely approximate the temporal pattern of each outcome indicator before the disruption occurrence, while the unweighted average sometimes fails. Both findings indicate the need for introducing causal inference framework to identify the true impact of disruptions.

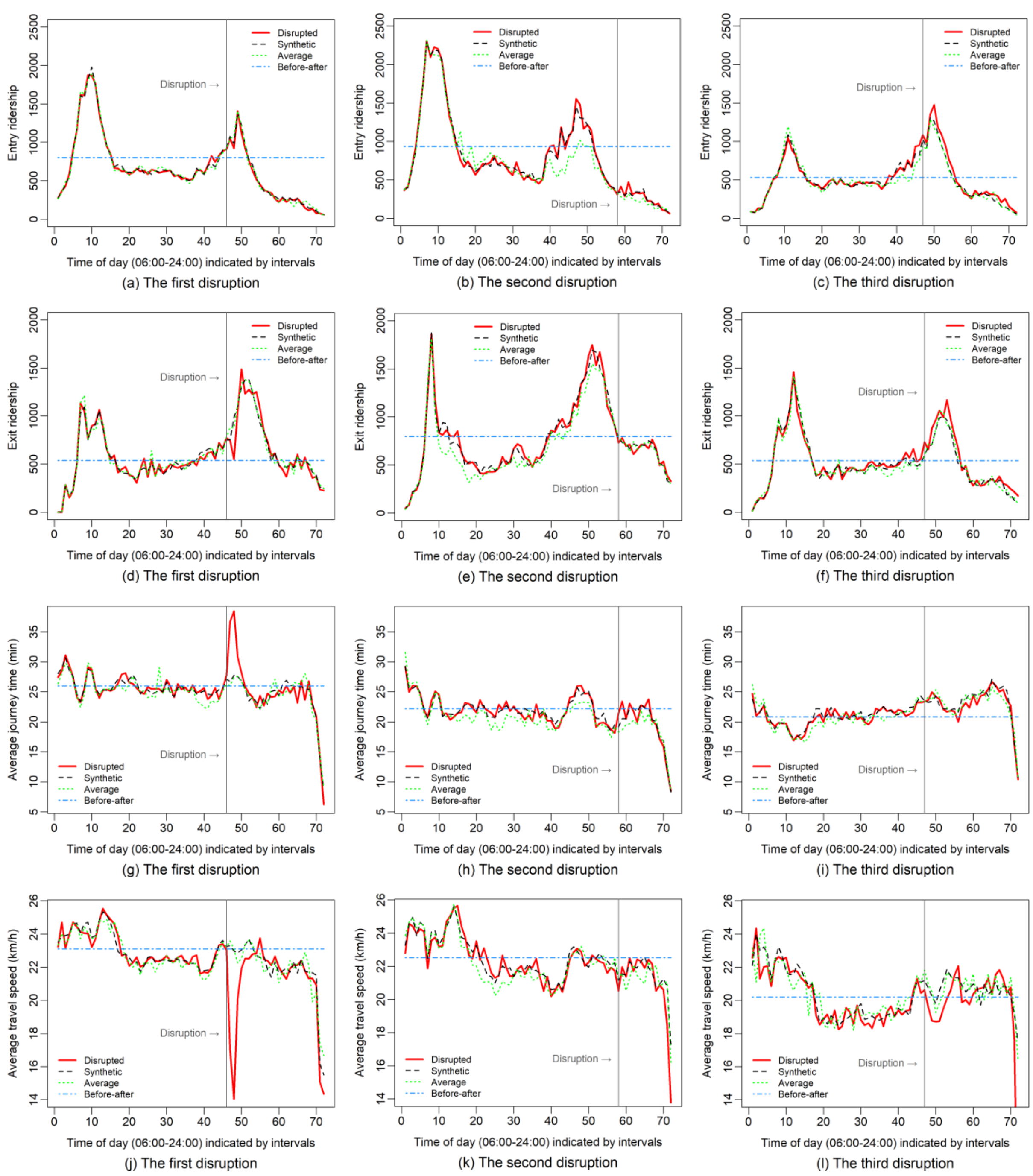

**Fig. 5.** Results of synthetic control estimation and direct causal effects on the three disruption scenarios – with comparison of two baseline methods.

By comparing the post-disruption patterns of the observed outcomes (red solid line) with their synthetical counterfactuals (black dashed line), we estimate the direct causal effects of the disruptions at three different stations, individually. Fig. 5 show that for the first example at a terminal station, where the interruption lasted 27 minutes, we observed notable decreases in exit ridership, indicating reduced demand following the disruption[7]. In contrast, the second disruption at a transfer station and the third disruption at a regular ‘through’ station exhibited only modest changes in passenger movements relative to the synthetic control reference, consistent with the shorter duration of interruptions at these locations. Furthermore, our analysis of passenger service levels at the first disruption reveals that passengers originating from this station experienced substantial increases in journey times (exceeding 11 minutes) and corresponding sharp declines in travel speed (up to 9 km/h), with these prolonged effects persisting even after the incident ended. Following a similar pattern, the minor disruption at the regular station resulted in comparatively smaller impacts on both travel times and speeds. During the moderate 12-minute disruption, however, the level of service was minimally impacted, likely because transfer stations offer passengers more alternative rerouting options.

[7] The variation in exit ridership could be driven by two countervailing mechanisms: (i) a concurrent reduction in alighting passengers due to the lack of incoming trains, and (ii) an increase of passengers who entered the station and then left after discovering the service suspension.

**Table 4.** Benchmarking - Mean square prediction errors of the five outcome variables, pre-treatment (first disruption).

| Method | Mean square prediction error (±S.E.*) | | | | |
|---|---|---|---|---|---|
| | Entry ridership | Exit ridership | Ave journey time | Ave travel speed | Crowding density |
| Synthetic control (ours) | $844.869_{\pm 5.435}$ | $1844.911_{\pm 11.674}$ | $0.496_{\pm 0.002}$ | $0.038_{\pm 1.523e\text{-}04}$ | $2.177e\text{-}06_{\pm 6.256\ e\text{-}07}$ |
| Average control | $1874.413_{\pm 10.044}$ | $2356.370_{\pm 15.602}$ | $1.632_{\pm 0.005}$ | $0.140_{\pm 3.017e\text{-}04}$ | $2.809\ e\text{-}05_{\pm 1.185e\text{-}06}$ |
| Single control | $2436.478_{\pm 229.437}$ | $5171.478_{\pm 249.073}$ | $2.524_{\pm 0.014}$ | $0.270_{\pm 0.001}$ | $1.010e\text{-}04_{\pm 4.125e\text{-}06}$ |
| LR | $21745.192_{\pm 2975.866}$ | $23623.784_{\pm 983.180}$ | $2.070_{\pm 0.232}$ | $1.837_{\pm 0.401}$ | $0.109_{\pm 0.012}$ |
| SVM** | $14432.690_{\pm 876.275}$ | $26101.43_{\pm 3306.45}$ | $1.305_{\pm 0.191}$ | $0.463_{\pm 0.185}$ | $0.076_{\pm 0.017}$ |
| RF** | $3263.353_{\pm 1075.033}$ | $3440.504_{\pm 1236.219}$ | $0.743_{\pm 0.085}$ | $0.288_{\pm 0.207}$ | $0.016_{\pm 0.006}$ |
| XGBoost** | $3337.424_{\pm 1026.356}$ | $3643.169_{\pm 1058.418}$ | $0.665_{\pm 0.094}$ | $0.281_{\pm 0.196}$ | $0.014_{\pm 0.007}$ |

*Standard errors are estimated by a bootstrapping algorithm, which randomly resamples (with replacement) the non-disrupted dates of the donor pool 1000 times. **Hyperparameters are tuned by random search to optimise cross-validation performance.

Then, as shown in Table 4 (focusing on the first disruption), we further validate the effectiveness of our approach in approximating pre-treatment outcomes, through comparison with the baselines described above. Overall, the proposed synthetic control framework achieves the lowest mean squared prediction errors across all variables, outperforming both the typical-day comparisons and the machine learning algorithms tested, which suggests its superior ability to construct counterfactuals and estimate unbiased causal effects. Moreover, the relatively small standard errors highlight the robustness of these estimates. The machine learning models appear less effective at replicating pre-disruption patterns, likely due to the limited size of donor pool (the occurrence of a day entirely free of incidents is inherently uncommon to observe within metro systems), which poses a small-sample challenge and prevents these methods from fully leveraging their strengths.

We also compare the mean values of the predictors $\boldsymbol{X}$ during the pre-treatment period across different baselines. For example, Table 5 illustrates the predictor distributions of the speed outcome variable for the first disruption. An advantage of machine learning methods is their direct use of predictors from disrupted units to predict counterfactual outcomes. However, our findings indicate that the weighted synthetic control can still provide a rather accurate approximation of predictor values, particularly in comparison to unweighted averages and the single control.

**Table 5.** Benchmarking - Mean values of predictors for average travel speed, pre-treatment (first disruption).

<table>
<tr><th rowspan="2">Predictors*</th><th rowspan="2">Disrupted unit<br>$\bar{X}_1^{a_I}$</th><th rowspan="2">Synthetic control<br>$\bar{X}_0^{a_I} C^{a_I *}$</th><th rowspan="2">Average control<br>$1/_J \bar{X}_0^{a_I}$</th><th rowspan="2">Single control<br>$\bar{X}_4^{a_I}$</th><th>LR</th><th>SVM</th><th>RF</th><th>XGB</th></tr>
<tr><th colspan="4">$\bar{X}_1^{a_I}$</th></tr>
<tr><td>Entry ridership</td><td>796.956</td><td>795.012</td><td>794.309</td><td>809.311</td><td colspan="4">796.956</td></tr>
<tr><td>Exit ridership</td><td>532.089</td><td>532.551</td><td>527.815</td><td>537.822</td><td colspan="4">532.089</td></tr>
<tr><td>Ave speed (km/h)</td><td>23.040</td><td>23.035</td><td>23.034</td><td>22.947</td><td colspan="4">23.040</td></tr>
<tr><td>Day of week (dummy)</td><td>1</td><td>0.152</td><td>0.154</td><td>0</td><td colspan="4">1</td></tr>
<tr><td>Temperature (℃)</td><td>19.272</td><td>20.995</td><td>22.051</td><td>18.235</td><td colspan="4">19.272</td></tr>
<tr><td>Wind (km/h)</td><td>7.244</td><td>10.463</td><td>13.460</td><td>13.444</td><td colspan="4">7.244</td></tr>
<tr><td>Rain (mm)</td><td>0.133</td><td>0.126</td><td>0.087</td><td>0</td><td colspan="4">0.133</td></tr>
<tr><td>Mega-event (dummy)</td><td>0</td><td>0.421</td><td>0.612</td><td>0.822</td><td colspan="4">0</td></tr>
</table>

*Auxiliary variables for the development of synthetic control units.

### 5.2 Placebo tests

In-place and in-time placebo tests are conducted to assess the sensitivity of our framework [82]. Specifically, these tests verify whether the estimated disruption effect is truly driven by the actual service interruption, rather than chance, model misspecification, or other confounding factors.

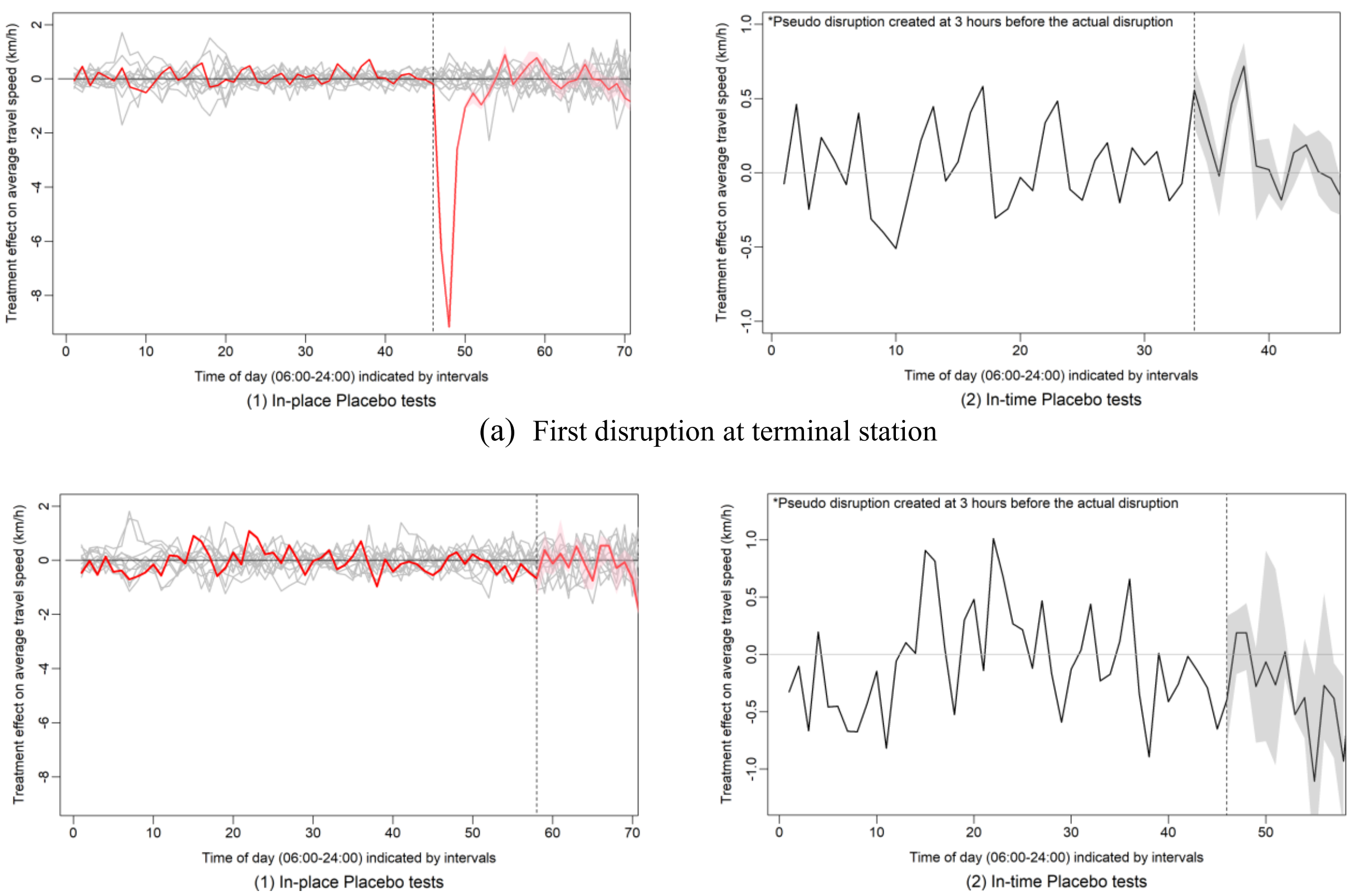

(a) First disruption at terminal station

(b) Second disruption at transfer station

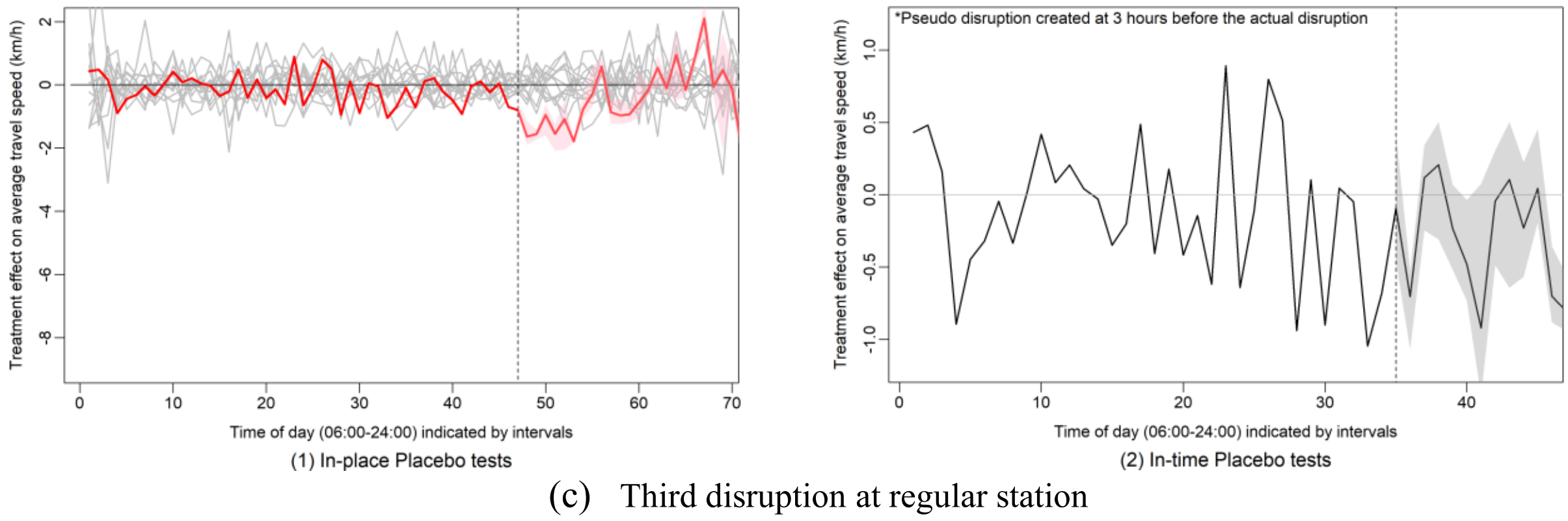

(c) Third disruption at regular station

**Fig. 6.** Results of in-place and in-time placebo tests for speed outcome variable. Falsification in time is created at 3 hours before the actual disruption.

We first perform the in-place placebo tests for average travel speed. For each example disruption, we randomly assign the "service interruption" to one of the non-disruptive days in the donor pool, and recompute the synthetic control using the remaining donor units. Under these pseudo-treated days, no sizeable change is expected in the outcome variables of interest. Fig. 6 (a1), (b1) and (c1) plot the difference between these hypothetical post-"treatment" paths (synthetic trend) for the 13 donor pool days, plus our main findings for the real disrupted days, depicted by the red line. We note that, for the first and third disruptions, the gaps estimated for the actual disruptions stand out from the distribution of placebo effects, consistent with our interpretation of Fig. 5.

Next, we carry out in-time placebo tests for average travel speed. For each example disruption, we set a "fake" disruption time three hours earlier than the actual incident and recompute the synthetic control using data only from the period before this fake timestamp. Fig. 6 (a2), (b2) and (c2) depicts the hypothetical post-"treatment" paths (still in the real pre-treatment period). We do not observe any significant changes in the speed outcome variable, indicating that our approach avoids spurious correlations and trend deviations unrelated to the actual intervention.

By combining these two types of placebo checks, we conclude that (i) our model is not inventing spurious impacts out of natural variability or flawed approximating, and (ii) the effect size found for the actual disruption is truly distinct from typical outcomes among the control units, ultimately confirming the robustness of the proposed synthetic control framework.

### 5.3 Spillover disruption effects and spatial-temporal propagation

Following the same manner that the synthetic control framework is applied to the disrupted station, we also estimate the causal effects for other non-disrupted (but nonetheless affected) stations, enabling us to capture the spatial and temporal propagation of impacts throughout the metro network. Using the first disruption as an example, we illustrate how its effects extend to the remaining 48 stations, particularly in terms of outcome variables such as average travel speed and crowding density.

Fig. 7 visualises the spatial progression of disruption impacts on average travel speed over five 15-minute intervals. The disruption initially occurred at the eastern terminus of the Island Line (marked by a star), primarily affecting trains traveling westbound. During the first 15 minutes, the station immediately following the terminus experiences severe delays (shown in red), with noticeable spillover effects spreading through several downstream stations. By the second 15-minute interval, these impacts continued westward along the line up to the tenth station, with the first four stations experiencing the highest level of delay. In parallel, the adjacent Tseung Kwan O Line in the northeast also shows signs of ripple effects, as several interchange and connecting stations begin to exhibit slower travel speeds.

Moving into the third and fourth intervals, although train services at the disrupted terminal station resumed at 18:15, delays continued propagating outward. While the upstream stations along the Island Line begin to gradually recover, many downstream stations and sections of the Tsuen Wan Line remain visibly slower. By 18:45, delay around the original disruption site continues to subside, but service degradation remains noticeable along the mid-sections of the Island line. Finally, by the fifth interval, roughly one and a half hours after the disruption, travel speeds had returned to near-normal at most stations. This temporal-spatial progression highlights how recovery often begins where the disruption initially occurred but can take longer to reach outlying segments of the line.

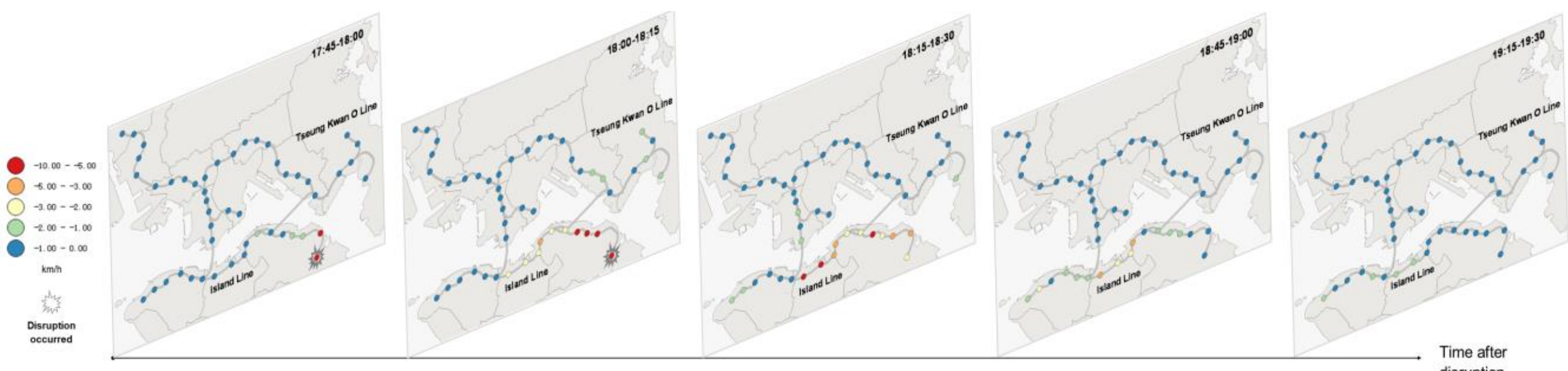

**Fig. 7.** Network propagation of disruption effects on average travel speed at different time periods. The star symbol denotes the location of the example disruption. Nodes represent metro stations, and their colour indicates the magnitude of speed reduction attributed to the disruption event.

Fig. 8 illustrates the spatiotemporal evolution of disruption-induced crowding on the Island Line, charting station-level passenger density (in-vehicle) across the line throughout the three-hour post disruption period. During evening peak hours, onboard crowding surged at most stations within an hour after the disruption occurred, then exhibiting considerable fluctuations. At certain high-demand, inner-city stations, the standing density exceeded 6 passengers per square metre, underscoring how disruptions even when originating at a remote terminus can propagate congestion far into the network core. Such extreme crowding not only increases passenger discomfort [85,86], but also prolongs station dwell times, potentially causing further delays downstream. Notably, we observe that transfer stations display lower peaks in crowding density changes compared to most regular stations on the Island Line, a pattern also reflected by the speed variations in Fig. 7. This evidence further suggests that transfer stations may be more resistant to disturbance from disruptions, likely due to additional route options, higher service frequencies, or greater overall capacity.

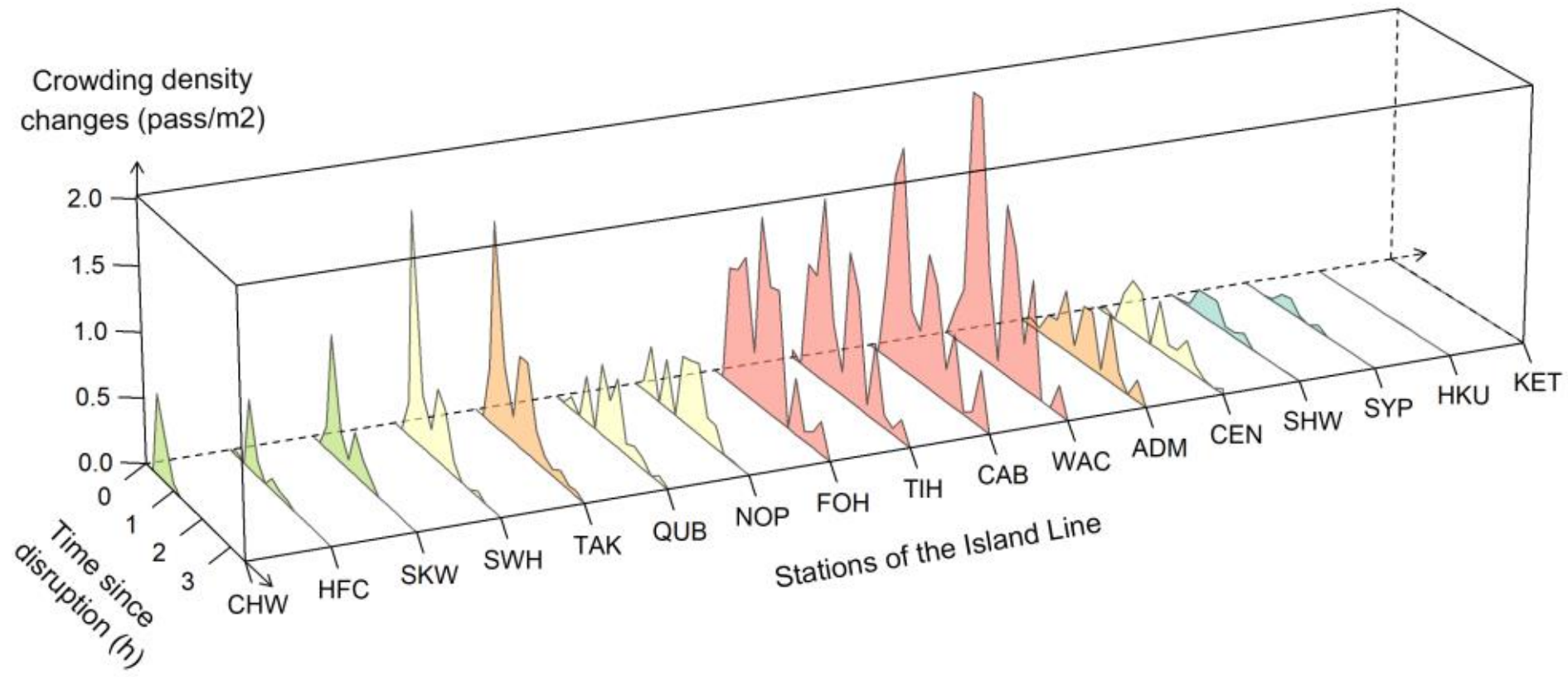

**Fig. 8.** Spillover effects on the in-vehicle crowding density (passengers/m2) at consecutive stations on the disrupted line. Colour coding represents the average level of crowding during and after the disruption: blue (0-0.5), green (0.5-1), yellow (1-2), orange (2-3), and red (3-6).

### 5.4 Causal resilience curves

Having estimated the direct and spillover causal effects of disruptions through our synthetic control framework, we next convert these impact time series into station-level causal resilience curves, illustrating how performance degrades and recovers in actual metro operations. Fig. 9 (a) to (c) demonstrate this transformation process using the first disruption as an example, presenting the temporal evolution of three key measures: (i) demand loss, (ii) passenger travel speed loss, and (iii) passenger comfort loss. Each panel highlights the different phases of the resilience lifecycle: pre-disruption baseline, disruption period, and post-disruption recovery.

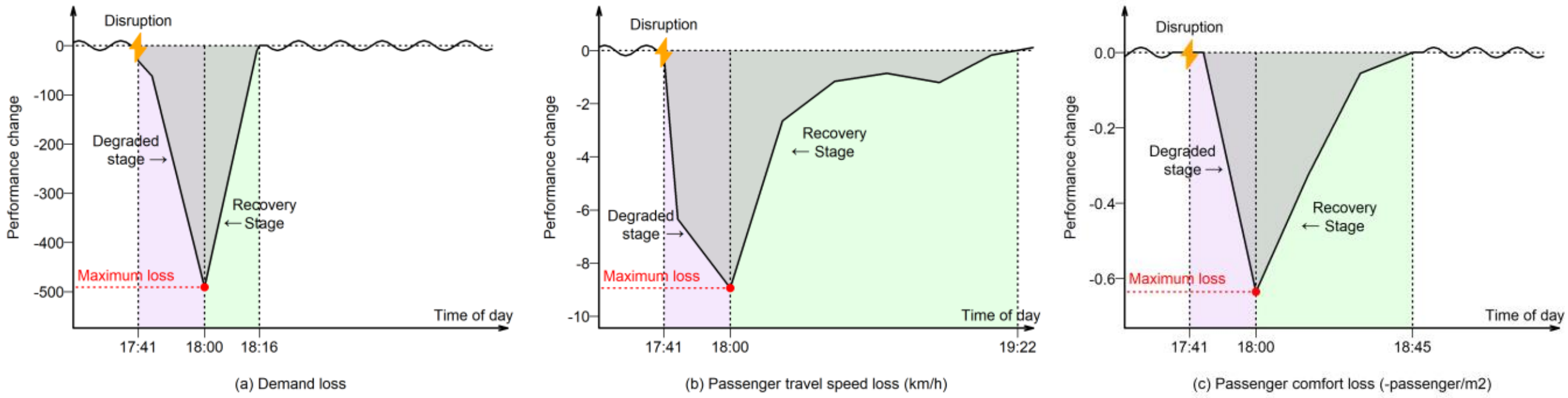

**Fig. 9.** Causal resilience curves of three performance measures for the first example disruption.

Fig. 10 extends this concept by displaying station-specific causal resilience curves across the affected network section. The stacked plots clearly demonstrate how the disruption's adverse effects radiate geographically and subside at varying rates, offering a detailed visualisation of the spatial-temporal patterns in performance degradation and subsequent recovery. Additionally, Fig. 8 can also be interpreted as a set of vertically inverted resilience curves for passenger comfort loss at stations along the Island Line.

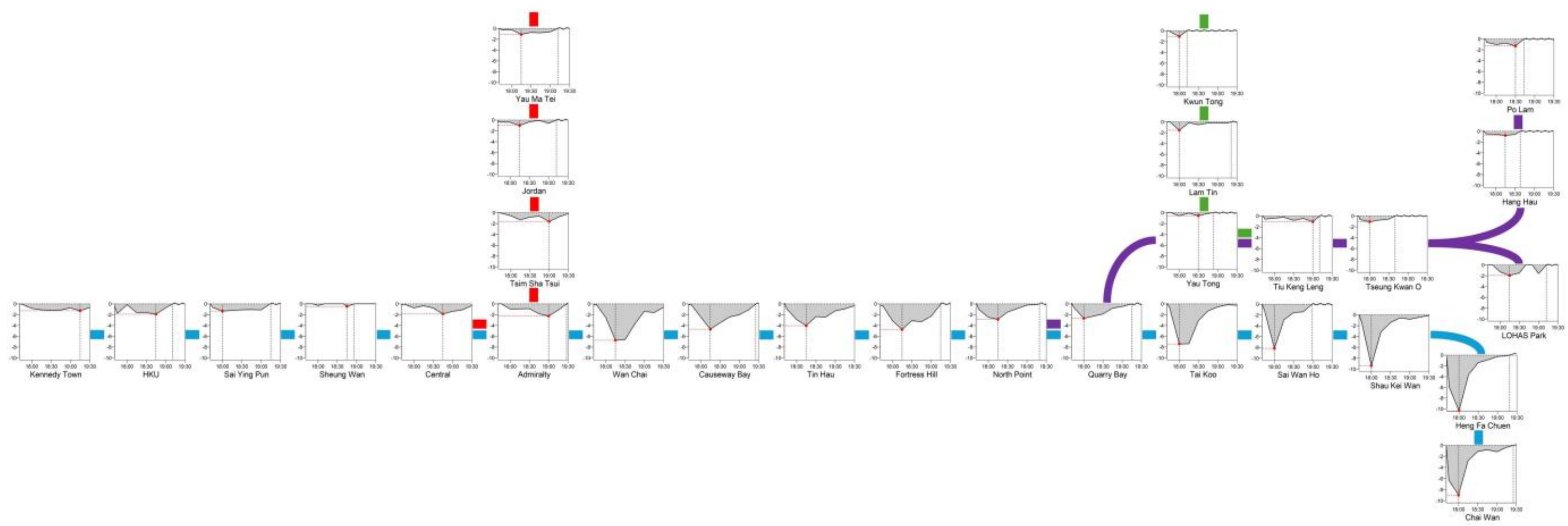

**Fig. 10.** Network-wide station-level causal resilience curves of passenger speed loss for the first example disruption.

By converting raw causal inference outputs into dynamic resilience curves, we translate complex impact information into actionable insights for operators and planners. These curves visually illustrate how quickly and severely a station's service performance declines, as well as how it recovers over time, which provide guidance for critical decisions such as scheduling additional services or issuing passenger advisories. They also enable a multidimensional resilience profile, allowing decision-makers to identify which aspects of service (e.g., crowding vs. speed) degrade the most and which recover the fastest, helping them prioritise interventions. Comparisons of these curves across stations expose systematic differences in resilience spatially, potentially guiding network redesign or targeted reinforcements in more vulnerable areas. Lastly, easily extracted metrics like maximum performance drop, time to recovery, or area of performance loss facilitate benchmarking and monitoring of disruptions over time, which helps operators assess the effectiveness of past recovery strategies. Altogether, this work delivers a novel empirical tool to help metro operators and researchers enhance daily operational planning and decision making.

Fig. 11 contrasts the resilience curves produced by the causal synthetic-control estimates (grey) with those obtained from the non-causal baselines (red-LR and blue-AC) introduced in Section 4. Because the average control method fails to adjust for confounding bias and the linear regression model lacks precision in undisrupted counterfactual prediction, these non-causal curves exhibit empirically implausible patterns. For instance, panels (a), (c) and (d) suggest an immediate rise in passenger demand and onboard comfort following the disruption occurrence, which is inconsistent with operational evidence. Furthermore, the AC and LR curves inaccurately characterise the duration and slope of both the degradation and recovery phases, and incorrectly place the turning point associated with maximum performance loss. In practice, reliance on such distorted curves can mislead resilience planning and improvement. Operators may underestimate the severity of the disruption (as illustrated in Fig. 11 (b), (d) and (e)) and consequently implement emergency measures informed by a spurious recovery profile, thereby wasting resources and diminishing the effectiveness of impact-mitigation efforts. These limitations highlight the necessity and importance of adopting causal resilience curves, which provide empirically robust, dynamically consistent, and unbiased representations of system performance under disruptions.

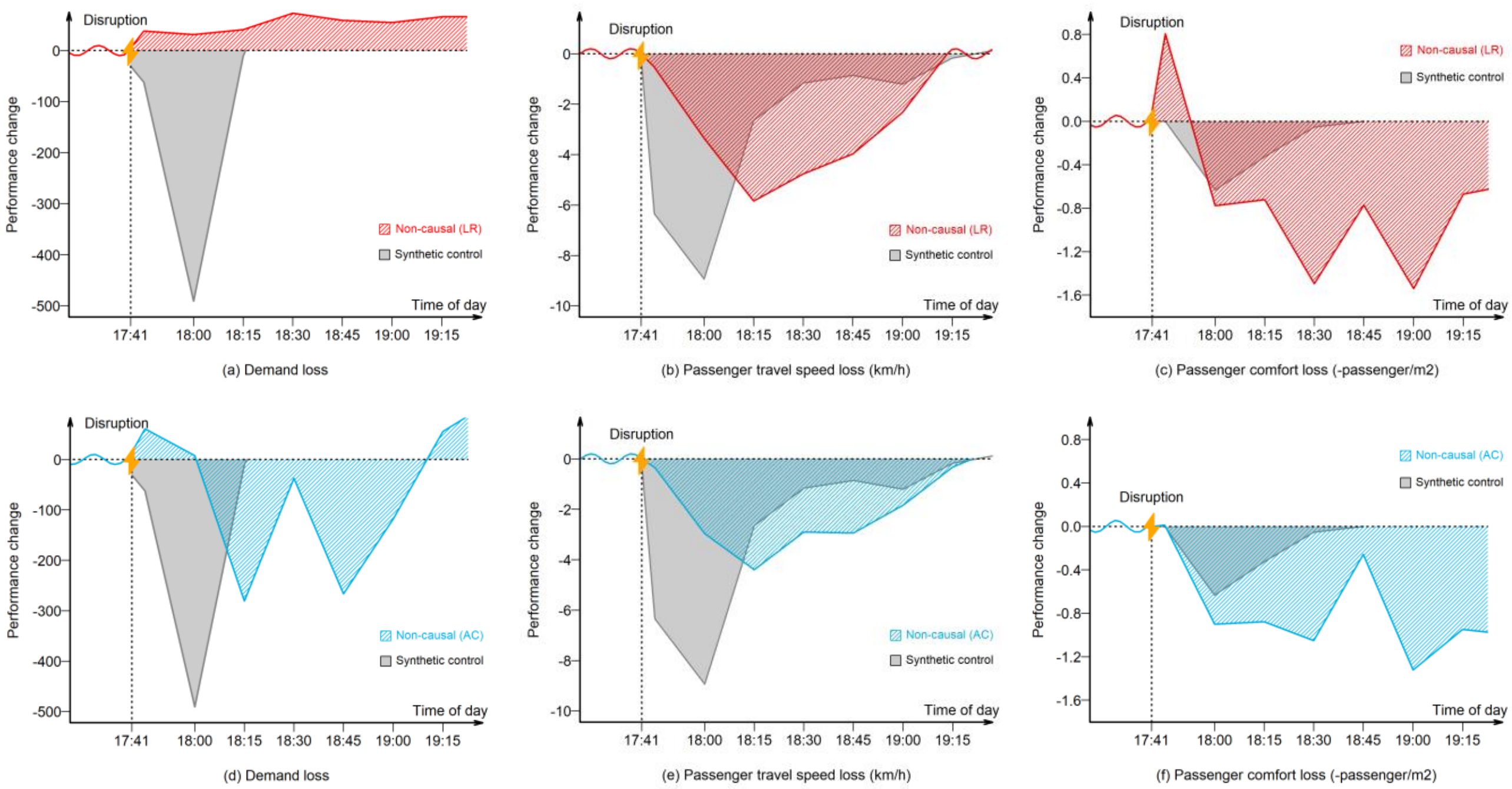

**Fig. 11.** Comparison with resilience curves derived using non-causal approaches: linear-regression prediction (LR) and average under normal conditions (AC).

### 5.5 Challenges and future work

While our results confirm the effectiveness of the proposed synthetic control framework, its applicability is subject to certain constraints. In particular, the data requirements of the approach remain relatively high, and meeting these requirements may pose practical challenges.

First, the construction of a suitable donor pool requires an adequate number of days with no disruptions in the entire metro system, which necessitates a sufficiently long data collection period, typically at least one month. If this condition is not met, the lack of sufficient donor-pool units can bias estimation results, as the method relies on unaffected observations to approximate counterfactual outcomes. As demonstrated by the standard errors in Table 4 and the confidence intervals in Fig. 6, adjusting the donor-pool composition through random sampling can introduce a degree of uncertainty into the disruption impact estimates. For larger and aging networks, where a completely disruption-free day is hard to find, a possible workaround is to use the ridership distribution OD matrix to verify that spillovers between distant sections are minimal, and then analyse those regions separately.

Second, the credibility of synthetic control estimators partially depends on having sufficient pre-treatment data [79]. Consequently, if a disruption occurs very early, such as within the first 15 minutes after the metro system opens, it lacks adequate pre-treatment information. In these cases, a near or perfect fit for predictor values may be spuriously achieved, undermining the robustness of the impact estimation. Developing alternative solutions for early-morning disruptions or situations with limited pre-treatment periods therefore becomes necessary.

Lastly, when the goal is to analyse individual disruption impacts, our method is best suited to days featuring either a single incident or multiple incidents whose effects do not overlap (i.e., the network fully recovers from one incident before another begins). However, in metro systems, it is not rare for multiple disruptions to occur on the same day with overlapping influences. In such instances, our framework will estimate their combined effect and cannot easily distinguish individual impacts, except during the interval between the occurrence of the first incident and the start of the next. This challenge of multiple concurrent treatments remains an important issue for synthetic control methods and causal inference more broadly, especially when accounting for network interference. Future research could focus on disentangling concurrent disruptions, exploring strategies to isolate each incident's contribution, and thus improve the granularity and applicability of the current framework.

Beyond these practical considerations, the proposed synthetic control framework is inherently scalable in both spatial and temporal dimensions, making it suitable for larger applications to larger metro networks and extended observation periods. Since the method relies on constructing a donor pool exclusively from days without disruptions, as well as independent calculation of synthetic controls for each station, its core principles remain intact even when the system size grows[8]. Furthermore, this framework can be adapted to assess disruptions and spillover effects in other public transport modes, such as commuter rail, trams, or bus rapid transit (BRT) systems, by tailoring the definition of disruptions to match each mode's operational characteristics. While some calibration may be required to account for differences in network topology, data formats, and service frequency, the ability to create synthetic control units for each station-interval pair remains the same. Consequently, our approach has the potential to become a generalisable tool for evaluating, benchmarking, and improving operational resilience across wider urban transit services.

## 6. Conclusions

Urban metros are instrumental in fostering sustainable mobility. However, service disruptions pose various challenges for metro systems by causing delays, overcrowding, and a drop in overall service quality. To address these challenges, this study introduced a customised synthetic control framework that transforms rich automated data into unbiased estimates of disruption impacts and empirically grounded causal resilience curves. Applied to a case study of Hong Kong MTR, the method proved superior in three aspects. First, the proposed causal inference framework outperformed traditional before-after and normal-day comparisons as well as advanced machine learning predictors in reproducing unbiased counterfactuals. Second, by quantifying the propagation of disruption spillovers, we uncovered pronounced spatial heterogeneity in resilience evolution patterns at station level. Terminal station failures generated protracted delays and crowding far beyond the incident site, whereas transfer stations exhibited lower performance losses owing to the availability of rerouting options. Lastly, the resulting causal resilience curves provide intuitive visualisations that translate multidimensional disruption impacts into actionable information for operators and planners.

The empirical evidence in this paper confirms that, when ignored, confounding factors and network spillover effects can severely bias resilience assessment. By constructing credible counterfactuals and tracing disruption propagation in space and time, the proposed framework equips agencies with a rigorous tool for prioritising infrastructure maintenance, optimising recovery strategies, and enhancing real-time passenger information provision. Looking ahead, our work can be extended to other public transport networks, such as bus rapid transit, heavy rail, or multimodal systems.

[8] Except that as the network grows over larger, the likelihood of observing a disruption-free day may diminish accordingly.

## 7. Acknowledgement

The authors are grateful for the support of the Hong Kong MTR, the data provider of this research. Any opinions, findings, and conclusions or recommendations expressed in this material are those of the authors and do not necessarily reflect the views of the MTR. Prateek Bansal was supported by the Leverhulme Trust Early Career Fellowship.

# Appendix

### A.1 Source of disruption data

Based on the detection method proposed by Zhang et al. [1], we transform the abnormal headway series that are extracted from the AVL data (train movements) into historical disruption data, which is then combined with official incident logs to build an accurate database of service disruptions. All records include the information of time and location of disruption occurrence, duration time and primary/secondary types.

Minor disruptions that lasted less than five minutes are excluded from the impact estimation. During the study period, 106 disruptions (of over 5 minutes) were observed on the four urban lines. Considering a primary disruption can spread along metro lines and lead to service interruption at other stations (secondary disruptions), the impacts of these two types of disruptions will be superimposed on each other and hence will be virtually indistinguishable. Thus, the causal effects estimated via the synthetic control framework are the integrated impacts from both the primary disruption and its corresponding secondary disruptions.

### A.2 Definition and calculation of outcome measures

*Entry ridership*: the number of passengers who enter the given station $a$, on day $d$, during the 15-minute interval $t$. This measure is calculated based on the tap-in records from the smart card data.

*Exit ridership*: the number of passengers who exit the given station $a$, on day $d$, during the 15-minute interval $t$. This measure is calculated based on the tap-out records from the smart card data.

*Average journey time*: the average of journey time of passengers who start their trips from the given station $a$, on day $d$, during the 15-minute interval $t$. This measure is calculated according to the timestamp of the paired tap-in and tap-out records.

*Average travel speed*: the average of the speed of all trips that start from the given station $a$, on day $d$, during the 15-minute interval $t$. For each trip, speed is computed as travel distance divided by observed journey time. Whereas journey time is directly obtained using the smart card data, travel distance (track length) of the most probable route is derived using the shortest path algorithm. Passengers who left the system and used other transport modes to reach their final destination are not included in the computation of this metrics. If the origin station is entirely closed and no passenger can continue trips by metro, then the average speed will be zero. If the origin station is partially closed, this metrics reflects the average speed of passengers who remain in the system.

*Crowding density on board*: the number of standing passengers per square metre on trains that pass through the given station a, on day d, during the 15-minute interval t. The calculation of this measure follows the method proposed by Hörcher et al. [2]. By merging smart card data with train movement data, passenger to train assignments are conducted to obtain the number of passengers on board each train. Then the crowding density equals the number of passengers on board subtracting the number of seats and dividing by the available floor area.